\documentclass[journal,transmag]{IEEEtran}
\usepackage{multirow}
\usepackage{romannum}
\usepackage{color}
\usepackage{graphicx}
\usepackage{dcolumn}
\usepackage{bm}
\usepackage{siunitx}
\usepackage[normalem]{ulem}
\usepackage{upgreek}

\begin{document}

\title{Phononic graded meta-MEMS for elastic wave amplification and filtering}

\author{\IEEEauthorblockN{Federico Maspero\IEEEauthorrefmark{3},
Jacopo Maria De Ponti\IEEEauthorrefmark{1},
Luca Iorio\IEEEauthorrefmark{1}, 
Annachiara Esposito\IEEEauthorrefmark{2},
Riccardo Bertacco\IEEEauthorrefmark{3},\\
Andrea di Matteo\IEEEauthorrefmark{2},
Alberto Corigliano\IEEEauthorrefmark{1}, and
Raffaele Ardito\IEEEauthorrefmark{1}}
\IEEEauthorblockA{\IEEEauthorrefmark{1} Dept. of Civil and Environmental Engineering, Politecnico di Milano, Piazza Leonardo da Vinci, 32, 20133 Milano, Italy}
\IEEEauthorblockA{\IEEEauthorrefmark{2}STMicroelectronics, Viale Remo De Feo, 1, 80022 Arzano, Italy}
\IEEEauthorblockA{\IEEEauthorrefmark{3}Dept. of Physics, Politecnico di Milano, Piazza Leonardo da Vinci, 32, 20133 Milano, Italy}
\thanks{Manuscript received June, 2023

Corresponding author: A. Corigliano (email: alberto.corigliano@polimi.it)}}

\IEEEtitleabstractindextext{%
\begin{abstract}

Inspired by recent graded metamaterials designs, we create phononic arrays of micro-resonators for frequency signal amplification and wave filtering. Leveraging suspended waveguides on a thick silicon substrate, we hybridize surface Rayleigh and Lamb flexural waves to effectively achieve phononic signal control along predefined channels. The guided waves are then spatially controlled using a suitable grading of the micro-resonators, which provide high signal-to-noise ratio and simultaneously create phononic delay-lines. The proposed device can be used for sensing, wave filtering or energy harvesting.
\end{abstract}
}

 \maketitle

\section{Introduction}


Over the past few decades the use of MEMS resonators has become increasingly important in the creation of new designs for timing references, RF filters, mass and motion sensors \cite{Corigliano2017}. Their properties ultimately depend on the ability to achieve a high quality ($Q$) factor combined with a high electromechanical coupling $\kappa_t^2$ \cite{Frangi2015AnchorLoss}. While previous technologies were mainly restricted by material limits, recent metamaterial solutions enable unprecedented high $Q$ factor and wide tuning range \cite{Cassella2020Periodic,Cassella2022Periodic}. Elastic and acoustic metamaterials have triggered intense research activity for their potential in wave control and manipulation \cite{Craster2013}, bringing new benefits to the world of MEMS and Radio Frequency (RF) signal control. This trend in exporting these concepts to the microscale can be seen from the recent amount of works in the field of metamaterials and phononic crystals (PnC) applied to MEMS. Sub-wavelength control and filtering of propagating surface acoustic waves has been achieved using annular holes and pillars \cite{Ash2017, Pouya2021, Ash2021} fully integrated into SAW devices \cite{ReviewSAW2019}. Micro-PnC have been employed to reduce anchor losses and consequently enhance $Q$ factor in MEMS resonators \cite{Ardito2016}, or to provide isolation from external vibrations \cite{Zega2020}. Defect-based waveguides have been adopted \cite{Zega2022} to efficiently couple distant MEMS resonators. Delay-lines programmed to selectively operate with large bandwidths around different frequencies \cite{Cassella2022DelayLine_a, Cassella2022DelayLine_b} have been designed by leveraging the unique dispersion features of metamaterial structures. While well-developed solutions for wave filtering or focusing have already been transferred to the micro scale \cite{Zega2020, GRINLensMicro2016}, other phenomena e.g.  rainbow reflection and trapping \cite{Colombi2016, DePonti2019, Chaplain2020Delineating}, non-local coupling \cite{roto_Wag}, space-time-modulated structures \cite{Riva-time}, mode conversion \cite{DePonti2021Selective, Pipe2022}, and topologically protected states \cite{Khanikaev2015, Mousavi2015, Huber2016, ChaplainTopological2020, PhysRevApplied.19.034079}, are still at inception.

\begin{figure}[h!]
\centering
\includegraphics[width = 0.5\textwidth]{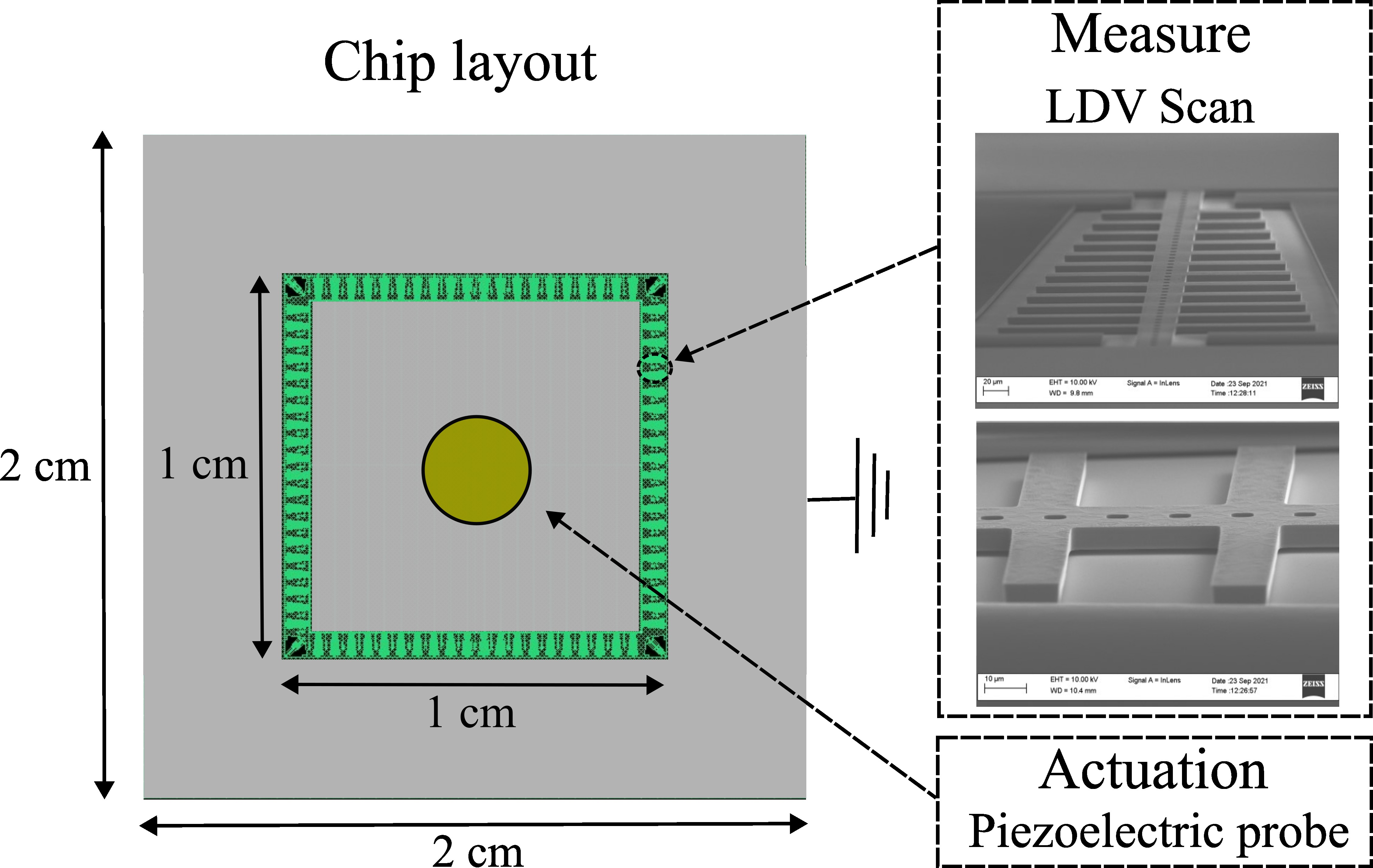}
\caption{Schematic of the device and the used experimental setup. In grey, the bulk silicon die is reported, where a piezoelectric probe is positioned at the center for actuation in its $d_{33}$ mode. This probe generates surface waves that arrive at the phononic graded meta-MEMS, depicted in green. The waves, while traversing the suspended arrays, slow down and the wave energy is confined in the lateral resonators, generating high displacement fields. A laser doppler vibrometer (LDV) evaluates the displacement field in the graded meta-MEMS.}
\label{schematic}
\end{figure}

In this paper we implement, at the microscale, a graded metamaterial configuration that is able to control elastic wave propagation via the so-called \emph{rainbow effect} \cite{DePonti2021b}. 
The challenge of this work is to develop, differently with respect to axial resonator solutions in RF filters \cite{Ash2017, Pouya2021, Ash2021}, an array of suspended flexural resonators, able to simultaneously provide wave filtering, amplification and signal delay. The proposed device then acts as a phononic waveguide, where the rainbow effect manifests its basic features, namely the confinement of different frequency components and the amplification along different spatial position \cite{Tsakmakidis2007}.
The interplay between these effects, namely filtering, amplification and signal delay, could open promising avenues in MEMS towards the development of highly efficient energy harvesters, in case the resonators are equipped with piezoelectric materials that may convert the elastic energy into electrical energy: the spatial confinement and amplification of elastic waves can greatly enhance the energy conversion. Moreover, in a long term perspective, the miniaturization of the device up to the nano-scale (that is feasible with up-to-date production processes) could increase the frequency range of the device in order to obtain ultra-wide band (UWB) filters for sensing and for next-generation 6G solutions.

The paper is organised as follows. In Sec. II, we introduce the device in terms of geometry and operating frequency range. In Sec. III, we provide a description of the fabrication methods and the experimental setup. In Sec. IV we compare numerical and experimental results, analysing the peculiar features of the device, i.e. the capability of slowing down waves, wave filtering and amplification. Finally, in Sec. V, we draw together some concluding remarks.

\section{Design of the device}

The phononic graded meta-MEMS consists of an elastic waveguide having a graded array of resonators of different lengths (see Figure \ref{schematic} and Figure \ref{fabrication}) attached to it. The term \textit{graded} refers to a smooth variation of a particular parameter of the local resonators along space (conventionally the resonance frequency), which enables spatially varying effective properties of the medium. 
The device is different with respect to other graded structures, such as GRIN lenses \cite{GRINLensMicro2016}, since it takes advantage of local band gaps to control wave propagation: guided waves slow down as they traverse the array with different frequency components localising at specific spatial positions. This phenomenon, known as the rainbow effect, has been observed in different realms of physics, from electromagnetism \cite{Tsakmakidis2007} to acoustics \cite{Zhu2013} and elasticity \cite{Colombi2016}, with multiple realizations for trapping \cite{Chaplain2020Delineating}, mode conversion \cite{DePonti2021Selective} or energy harvesting \cite{DePonti2021Book, Zhao2022}. 

In this work, we consider low frequency resonators (less than 100 MHz)  - commonly used in low-precision consumer electronics - given the constraints on the prototype dimensions dictated by the manufacturing process and the experimental setup. Due to the intrinsic limitations of the micro-fabrication and the problem in generating the input elastic waves at high frequencies, we opt for a device working in the frequency range between 200 kHz and 1.2 MHz. It is worth noting that the proof-of-concept, obtained at the scale of the prototype, can be usefully exploited for the application at smaller scales, provided that a suitable production process is adopted.

A detailed description of the device size and features is reported in Tab. \ref{Table1}.

\begin{table}[h!]
\centering
\caption{Graded meta-MEMS geometry}
\begin{tabular}{ |c|c| } 
 \hline
 Parameter &  Value \\ 
 \hline
 Number of resonators & 12  \\  Resonators width & 15 [\textmu  m]  \\ 
 Minimum length of resonator & 50 [\textmu  m]  \\ 
 Maximum length of resonator & 110 [\textmu  m]  \\
 Waveguide length & 750 [\textmu  m]  \\ 
 Waveguide width & 30 [\textmu  m]  \\ 
 Thickness of the suspended parts & 5 [\textmu  m]  \\
 Thickness of the central bulk part & 730 [\textmu  m]  \\
 \hline
\end{tabular}
\label{Table1}
\end{table}

\section {Device fabrication and characterisation} 

\subsection{Device fabrication}

\begin{figure}[t!]
    \centering
    \includegraphics[width = 0.5\textwidth]{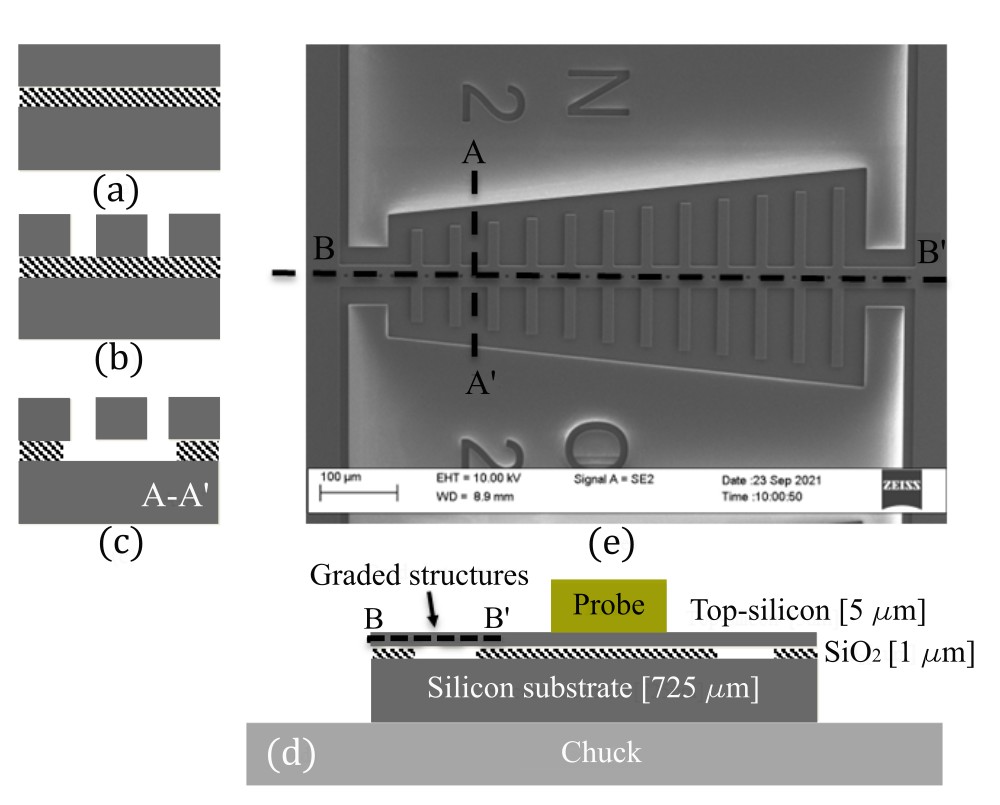}
    \caption{(a)-(c) Schematic process flow of the fabrication process. (d) Detail view of the device cross section. (e) Scanning electron microscope image of the fabricated device.}
    \label{fabrication}
\end{figure}

The samples are fabricated starting from a Silicon On Insulator (SOI) wafer having 5 \textmu m of n-doped silicon on top of 1 \textmu m of silicon oxide (Figure \ref{fabrication}a). The substrate size is 2x2 cm (Figure \ref{schematic}), its overall thickness is approximately 730 \textmu m and it is obtained by dicing an 8 inches wafer.

The litography is performed with a maskless aligner, model MLA100 from Heidelber instrument. After the litography, the MEMS structures are defined using a DRIE Bosch-like process \cite{laermer_method_1996} performed with a reactive ion etching, model Oxford Plasmalab 100 (Figure \ref{fabrication}b). Finally, the devices are released by wet etching of the silicon oxide layer in buffered oxide etch (Figure \ref{fabrication}c). A global cross-section of the device is depicted in Figure \ref{fabrication}b, that  shows the difference between the suspended parts and the bulky central region, and an image of one of the fabricated device is reported in Figure \ref{fabrication}e.
This procedure allows us to obtain several suspended microstructures, all aligned along a squared perimeter with a side of 1 cm, to accommodate the probe in the center, able to actuate Rayleigh surface waves (Figure \ref{schematic}).

\subsection {Measurement setup and procedure}

\begin{figure}[b!]
    \centering
    \includegraphics[width = 0.48\textwidth]{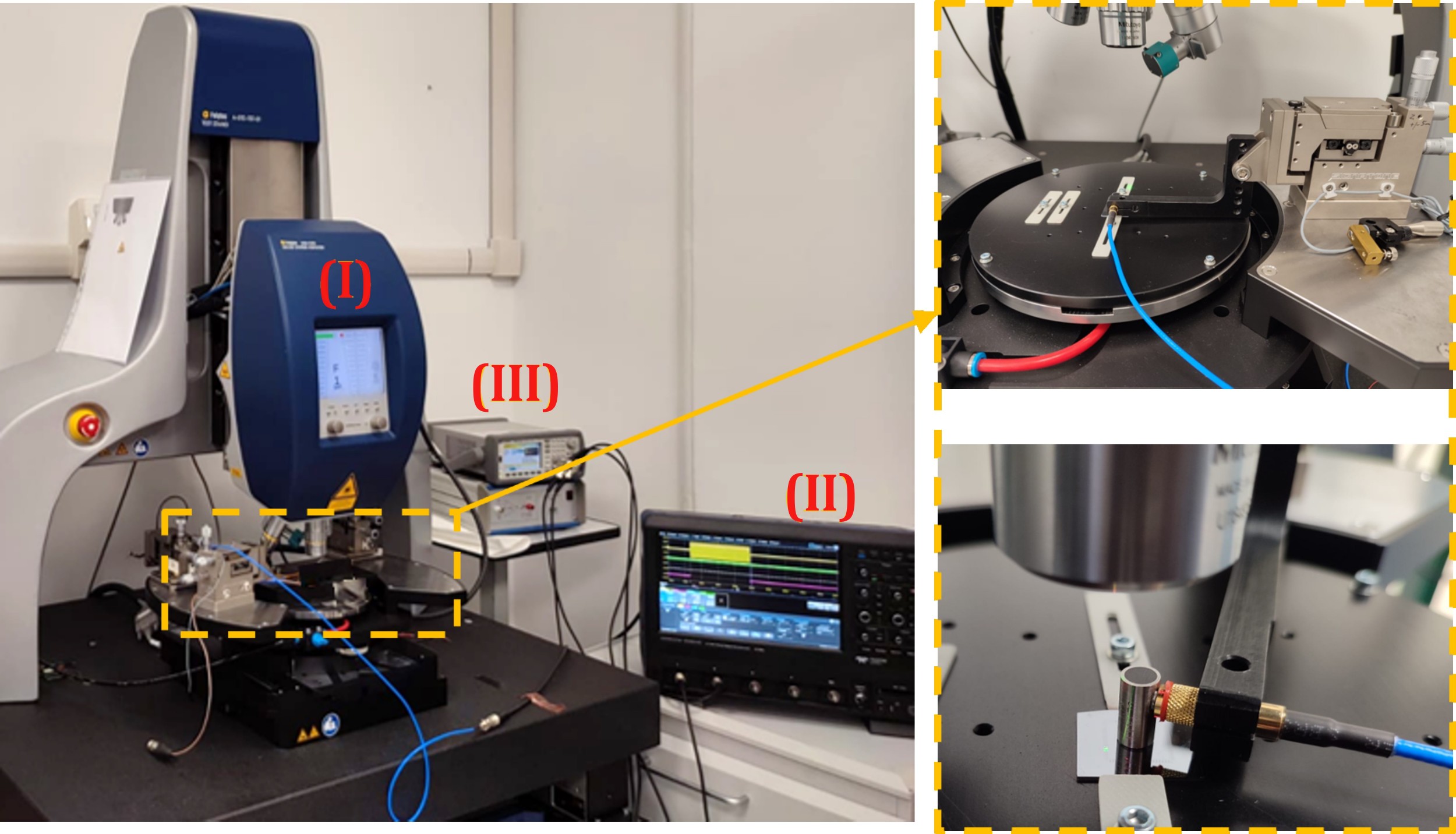}
    \caption{Experimental setup for the characterization of the microscale structure. (a) The wavefield in space and time is measured using a Polytec MSA500 laser doppler vibrometer (I), eventually complemented with an oscilloscope (II), while the input is created using a signal generator (III). The chip is placed on the thick chuck of the vibrometer and is forced in the center through a probe.}
    \label{setup}
\end{figure}

The measurement setup is depicted in Figure \ref{setup}. The silicon chip is placed on the chuck of a probe station underneath a vibrometer system of type Polytec MSA500.

The system can detect the out-of-plane displacements in the picometer range with an in-plane position accuracy of the laser beam limited by the beam spot size ($\approx$ 0.9 \textmu m for the 50X magnification); the frequency range of detection of the instrument is 20 MHz. The system can acquire a grid of points defined by the user and reconstruct a vibration map of the device. 
The mechanical input is provided by means of a piezoelectric probe of type Ultran, series KC12-1, which is kept in direct contact with the device under test by a rigid arm. The probe size was chosen as small as possible in order to fit on a relatively small silicon substrate and the frequency range of excitation delivered by the probe set the frequency boundaries of our experiment i.e. from 200 kHz to 1.2 MHz. The probe is driven by a power amplifier in sync to the MSA500 or to an external arbitrary wave-form generator. 

Two types of measurements are performed. First a wide bandwidth sweep signal with a total amplitude of 3 V is generated to excite all the frequencies ranging from 300 kHz to 1300 kHz. The steady-state response of the vibrational map is acquired, providing the frequency spectrum of the displacement for each point of the grid. This analysis is done to obtain the frequency response function (FRF) of the fabricated device, to verify the agreement between numerical predictions and experimental results. Numerical calculations are performed through ABAQUS \textsuperscript{\textregistered} finite element software, for one suspended graded meta-MEMS on an infinite medium modeled using Absorbing Boundary Conditions \cite{art:Rajagopal12}. The frequency analysis is carried out in order to characterise the resonance frequency of each cantilever, which is a preliminary step for the subsequent analyses. The second measurement procedure consists in driving the piezoelectric probe with a burst sinusoidal signal generated by an external wave-form generators; this allows us to investigate the transient behaviour, detecting  wavenumber transformations or mode conversions. The oscillation induced by the burst is then tracked along the waveguide in order to characterise the signal in time and space.

\section{Results and discussion}

The device controls the propagating wave in the following ways: (i) it slows down the wave, lowering both the phase and the group velocity, (ii) it filters specific frequency components of the input and (iii) it amplifies the displacement and velocity fields of the lateral resonators induced by the incoming wave at their target resonant frequencies. From an application point of view, these features translate respectively into: (i) delay-lines, (ii) frequency filters and (iii) frequency selective amplifiers for surface waves.

In order to assure such phenomena, we orient the graded meta-MEMS so that the incoming wave, generated at the center, propagates from the shortest (first) to the longest (last) resonator. In this way, the wave propagation is accompanied by a slow wavenumber transformation due to a decrease in the resonant frequencies along space.

\begin{figure}[h!]
    \centering
    \includegraphics[width = 0.48\textwidth]{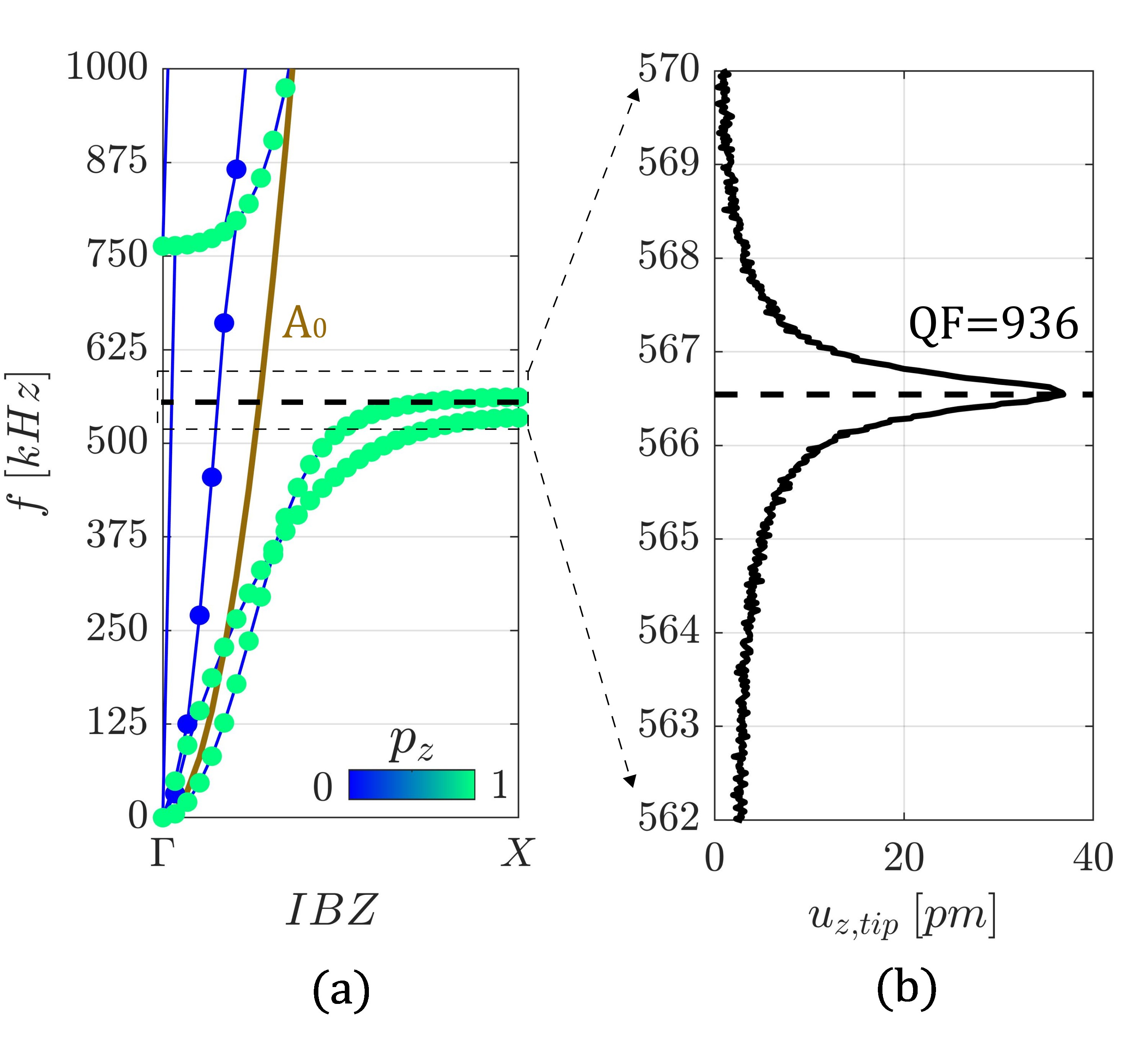}
    \caption{(a) Numerical dispersion relation for elastic waves propagating in the wave-guide. An infinite repetition of periodic resonators is assumed to properly detect the local dispersion relation in a given spatial position. The dashed line represents the theoretical frequency of the target resonator (third last). (b) Experimental measurement of the resonance peak of the target resonator.}
    \label{bandgap}
\end{figure}

The first result of the characterisation is reported in Figure \ref{bandgap}. 

\begin{figure*}[t!]
    \centering
    \includegraphics[width = 0.97\textwidth]{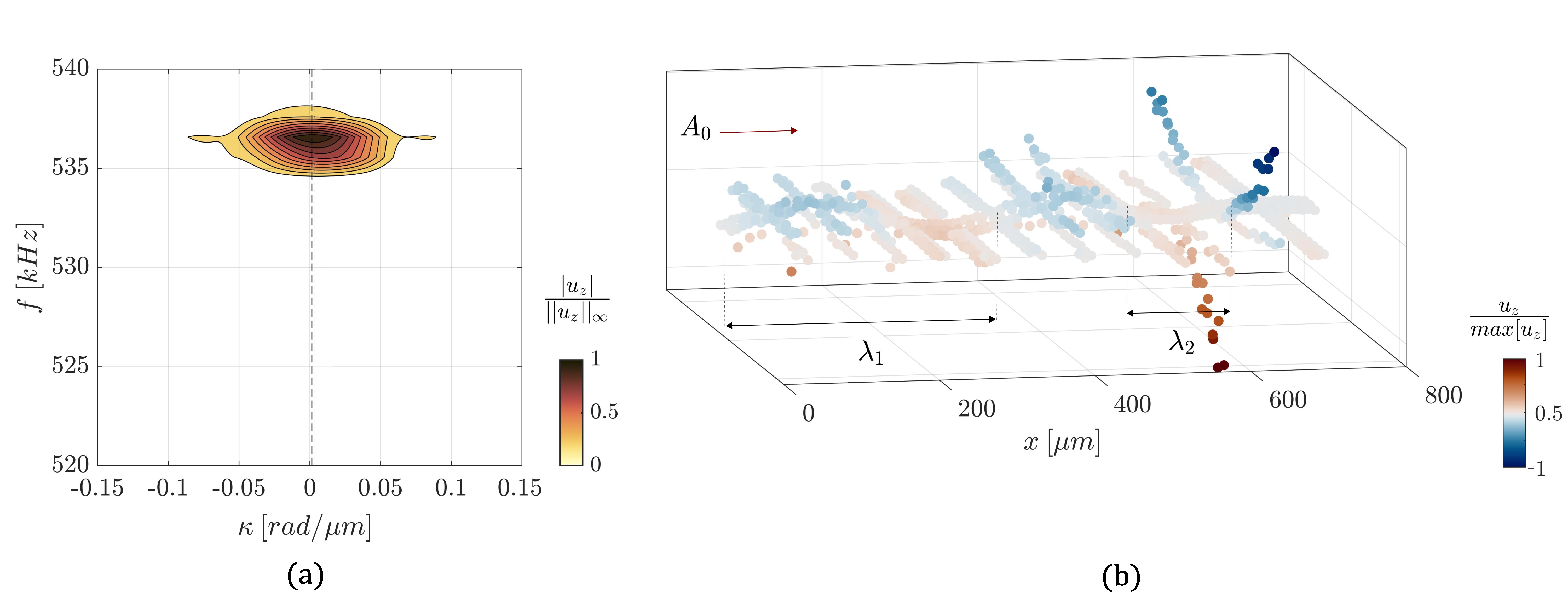}
    \caption{(a) Experimental results of the input wave generated with a burst at 536 kHz compared to the analytical dispersion relation of a Rayleigh wave (dashed line). (b) Displacement field of the rainbow array at a specific time interval during the analysis. A wavenumber transformation can be observed inside the structure.}
    \label{delay}
\end{figure*}

Provided the grading is gentle enough, the behaviour of the entire structure can be deduced from the knowledge of the local dispersion curves \cite{Colombi2016}. The dispersion relation displays hybrid modes which are obtained by coupling the dynamic properties of the plain waveguide and the resonators. Since we are considering flexural wave propagation in the suspended part, the reference wave mode $A_0$ couples with the flexural modes of the resonator, resulting in horizontal bands at their resonant frequencies; this allows us to detect the resonant frequencies of a given resonator from the zero group-velocity dispersion branches.
For this reason, we compare in Figure \ref{bandgap} the dispersion curve of a given resonator (the third last in this case) with the corresponding experimental Frequency Response Function (FRF) associated to the first flexural mode.

The numerical prediction of the resonance obtained from the dispersion curve and the experimental data, extracted from the Fast Fourier Transform (FFT) performed by the vibrometer over the entire structure, are in excellent agreement with a $\sim$ 0.9$\%$ relative error in the resonant frequency prediction.

This preliminary analysis shows that the micro-fabrication process is accurate and the tested structures are comparable to the initial design. Furthermore, to characterise the dynamic motion of the resonator and to assess the real damping, the experimental quality factor of the considered resonator is $Q := \omega_r/\Delta \omega_r = 936$, being $\omega_r$ the angular resonant frequency and $\Delta \omega_r$ the angular half-power bandwidth, both values being retrieved from the experimental data.  
After having verified that the obtained frequencies agree with the design target, further analyses are performed to characterise the propagating elastic wave on the chip and to assess the ability of the graded meta-MEMS to control, filter and confine the elastic signal that traverses the system. 

The first necessary step is to define the input signal generated by the piezoelectric probe. In general, a source excitation on a thick medium generates propagating surface and bulk shear/pressure waves \cite{Achenbach1973, Graff1991}. 
Assuming the medium, composed of the silicon substrate and the chuck, sufficiently thick with respect to the wavelength, we can  realistically infer the generation of Rayleigh and bulk waves. This is the type of wave that is needed for our structures to work properly, given that a Rayleigh wave is easily converted into an out-of-plane flexural mode when it encounters the beams that are the backbone of the graded meta-MEMS. To assess the input wave, a time burst analysis at 536 kHz is performed on a line of points on the chip before the waveguide. The signal is then converted in the frequency-wavenumber domain through a double FFT transform both in time and space. The goal is to correctly define the nature of the wave by comparing the experimental result with the analytical solution of the Rayleigh's dispersion curve. The result is reported in Figure \ref{delay}(a) where the dotted line reports the analytical solution of the dispersion relation of a Rayleigh wave \cite{Achenbach1973, Graff1991} for an infinite medium, while the contour plot represents the experimental solution. Even if the  wide experimental wavenumber content is a hallmark of several types of waves, the maximum of the spectrum is located at very low wavenumbers: this suggests long wavelengths ($\approx$ 9 mm), fully compatible with Rayleigh waves.

\subsection{Micro-mechanical delay line}

To observe the rainbow metamaterial capability of slowing down the traversing wave and to confine it, we perform a time domain analysis with a relatively narrow-band signal at 536 kHz. This frequency is slightly below the resonant frequency of the third last resonator, such that the wave is not stopped but strongly slowed down. We consider, as a reference, a resonator located at the end of the waveguide to fully show the resonators' ability of shortening the wavelength and simultaneously slowing down the wave. This effect happens more efficiently with frequency in the lower range of the array.
Figure \ref{delay} (b), shows a time frame from the experiments. We see how the incoming wave propagates through the waveguide and it is gradually compressed in space, given that we discern 2 distinct wavelengths $\lambda$. This compression is also matched by a slowing of both phase velocity (which is a result of shortening the wavelength at a given frequency) and group velocity (given by a change in shape and curvature of the dispersion curve). Moreover, the resonators are acting as expected, moving out-of-plane and confining the propagating wave inside them.

\subsection{Wave filtering and signal amplification}

\begin{figure*}[t!]
    \centering
    \includegraphics[width = 0.95\textwidth]{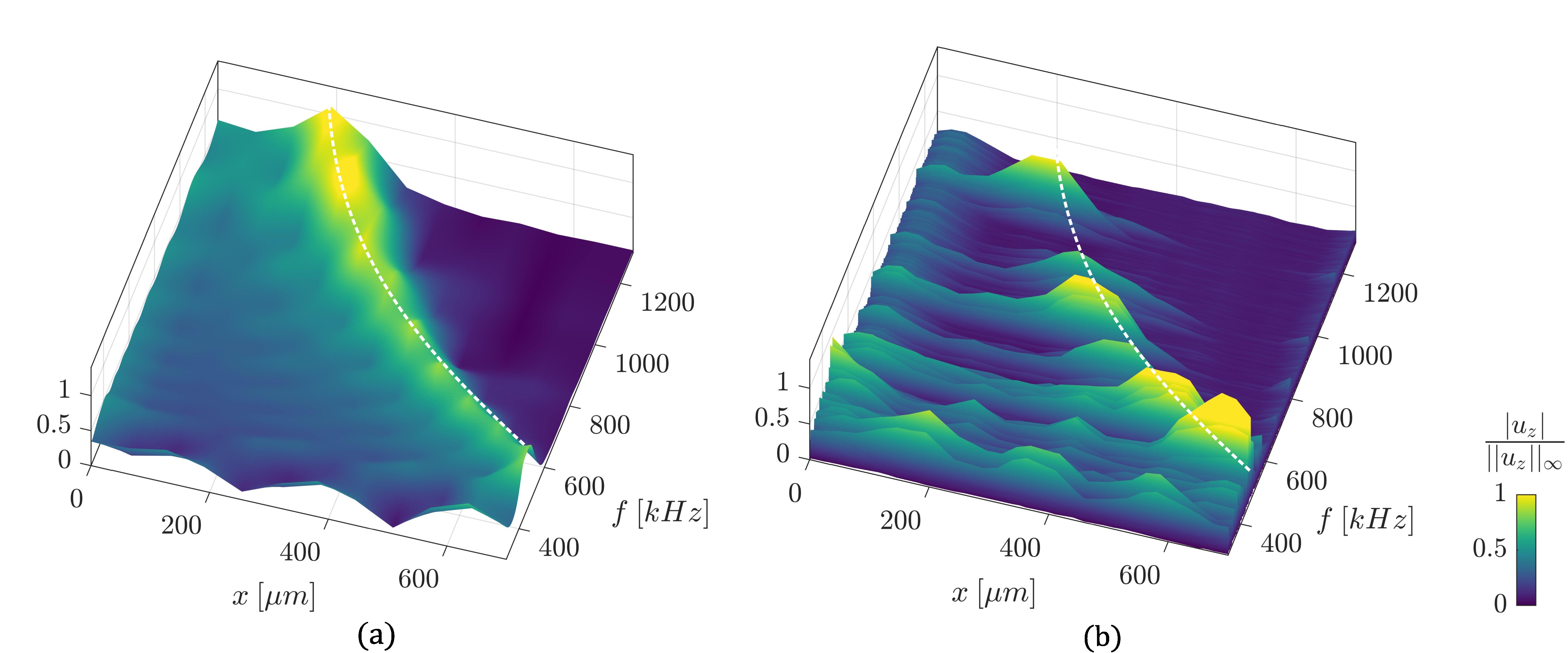}
    \caption{Space-frequency results of the numerical (a) and experimental (b) data of the out of plane displacement field. The white line, obtained by interpolating the local maxima, denotes the space-frequency dependence of the bandgap inside the graded meta-MEMS.}
    \label{rainbow}
\end{figure*}

The final characterisation aims to show the effectiveness of the graded meta-MEMS in acting as frequency filters and amplifiers. To verify this point, a large bandwidth signal is generated at the input.  Figure \ref{rainbow}, shows a space-frequency analysis of the numerical and experimental data. The different frequency components of the wave are selectively stopped at different spatial positions, according to a phenomenon known in physics as as the rainbow effect \cite{Tsakmakidis2007}. The halt is in  correspondence of the bandgap openings that are generated progressively in space by the resonance frequency of the resonators. Moreover, we notice that the displacement field of each wave component is increased in correspondence of the position of the resonator that has a resonance frequency matching the one of the wave. Once the maximum amplitude for a given frequency is reached, the bandgap opens and strong wave attenuation and reflection is observed. It is clearly visible how the frequency content of the input signal is amplified by the resonance of each resonator. The amplification at resonance is increased by the group velocity decrease, which enhances the interaction time between the wave and the resonators \cite{DePonti2019}. In this manner, before being backscattered, the waves enjoy a longer interaction with the resonators enhancing their displacement amplitude when compared against the case of isolated resonators with equal resonance frequency.

These two effects, i.e. filtering and amplification are strongly related; the wave attenuation along the waveguide is caused by the transfer of energy to the resonators, thus experiencing amplification effects.

\section{Conclusions}

In this paper we have shown the rainbow effect at the microscale by means of flexural resonators. We have experimentally demonstrated the potential advantages of using phononic graded meta-MEMS for frequency signal amplification and filtering. Leveraging a peculiar control of the waveguide dispersion properties using graded resonators we confine and amplify different frequency components along space. The implementation of locally changing dispersion curves integrates a slowdown effect both in terms of energy and information carried by the wave. This effect also allows for broadband signal attenuation given the generation of multiple partially overlapping band gaps. The proposed device can be suitably employed for applications involving sensing, wave filtering or delay-lines in MEMS. Moreover, it paves the way to the development of efficient energy harvesters, that may benefit from the enhanced interaction of the incoming wave with the target resonator.      

\section{Acknowledgements}
The authors acknowledge the  financial support of the H2020 FET-proactive Metamaterial Enabled Vibration Energy Harvesting (MetaVEH) project under Grant Agreement No. 952039. The authors acknowledge the availability of experimental facilities at PoliFAB.

\bibliographystyle{IEEEtran}

\bibliography{bibliography_micro_ant}

\begin{thebibliography}{10}
\providecommand{\url}[1]{#1}
\csname url@samestyle\endcsname
\providecommand{\newblock}{\relax}
\providecommand{\bibinfo}[2]{#2}
\providecommand{\BIBentrySTDinterwordspacing}{\spaceskip=0pt\relax}
\providecommand{\BIBentryALTinterwordstretchfactor}{4}
\providecommand{\BIBentryALTinterwordspacing}{\spaceskip=\fontdimen2\font plus
\BIBentryALTinterwordstretchfactor\fontdimen3\font minus
  \fontdimen4\font\relax}
\providecommand{\BIBforeignlanguage}[2]{{%
\expandafter\ifx\csname l@#1\endcsname\relax
\typeout{** WARNING: IEEEtran.bst: No hyphenation pattern has been}%
\typeout{** loaded for the language `#1'. Using the pattern for}%
\typeout{** the default language instead.}%
\else
\language=\csname l@#1\endcsname
\fi
#2}}
\providecommand{\BIBdecl}{\relax}
\BIBdecl

\bibitem{Corigliano2017}
A.~Corigliano, R.~Ardito, C.~Comi, A.~Frangi, A.~Ghisi, and S.~Mariani,
  \emph{Mechanics of Microsystems}.\hskip 1em plus 0.5em minus 0.4em\relax John
  Wiley '\&' Sons Inc, 2017.

\bibitem{Frangi2015AnchorLoss}
J.~Segovia-Fernandez, M.~Cremonesi, C.~Cassella, A.~Frangi, and G.~Piazza,
  ``Anchor losses in {AlN} contour mode resonators,'' \emph{Journal of
  Microelectromechanical Systems}, vol.~24, no.~2, pp. 265--275, 2015.

\bibitem{Cassella2020Periodic}
X.~Zhao, L.~Colombo, and C.~Cassella, ``Aluminum nitride
  two-dimensional-resonant-rods,'' \emph{Appl. Phys. Lett.}, vol. 116, no.
  143504, pp. 1--6, 2020.

\bibitem{Cassella2022Periodic}
X.~Zhao, O.~Kaya, M.~Pirro, M.~Assylbekova, L.~Colombo, P.~Simeoni, and
  C.~Cassella, ``A {5.3 GHz Al0.76Sc0.24N} two-dimensional resonant rods
  resonator with a kt2 of 23.9\%,'' \emph{Journal of Microelectromechanical
  Systems}, vol.~31, no.~4, pp. 561--570, 2022.

\bibitem{Craster2013}
R.~V. Craster and S.~Guenneau, \emph{Acoustic Metamaterials}.\hskip 1em plus
  0.5em minus 0.4em\relax Springer Dordrecht, 2016.

\bibitem{Ash2017}
\BIBentryALTinterwordspacing
B.~J. Ash, S.~R. Worsfold, P.~Vukusic, and G.~R. Nash, ``A highly attenuating
  and frequency tailorable annular hole phononic crystal for surface acoustic
  waves,'' \emph{Nature Communications}, vol.~8, no. 174, pp. 1--7, 2017.
  [Online]. Available: \url{https://www.nature.com/articles/s41467-017-00278-0}
\BIBentrySTDinterwordspacing

\bibitem{Pouya2021}
\BIBentryALTinterwordspacing
C.~Pouya and G.~R. Nash, ``Sub- and supersonic elastic waves in an annular hole
  phononic metamaterial,'' \emph{Communications Materials}, vol.~2, no.~55, pp.
  1--8, 2021. [Online]. Available:
  \url{https://www.nature.com/articles/s43246-021-00163-w}
\BIBentrySTDinterwordspacing

\bibitem{Ash2021}
\BIBentryALTinterwordspacing
B.~J. Ash, A.~R. Rezk, L.~Y. Yeo, and G.~R. Nash, ``Subwavelength confinement
  of propagating surface acoustic waves,'' \emph{Applied Physics Letters}, vol.
  118, no. 013502, pp. 1--7, 2021. [Online]. Available:
  \url{https://aip.scitation.org/doi/full/10.1063/5.0038381}
\BIBentrySTDinterwordspacing

\bibitem{ReviewSAW2019}
\BIBentryALTinterwordspacing
P.~Delsing~et al., ``The 2019 surface acoustic waves roadmap,'' \emph{Journal
  of Physics D: Applied Physics}, vol.~52, no.~35, pp. 1--41, 2019. [Online].
  Available:
  \url{https://iopscience.iop.org/article/10.1088/1361-6463/ab1b04/meta}
\BIBentrySTDinterwordspacing

\bibitem{Ardito2016}
R.~Ardito, M.~Cremonesi, L.~D'Alessandro, and A.~Frangi, ``Application of
  optimally-shaped phononic crystals to reduce anchor losses of {MEMS}
  resonators,'' in \emph{2016 IEEE International Ultrasonics Symposium (IUS)},
  2016, pp. 1--3.

\bibitem{Zega2020}
Z.~Yao, V.~Zega, Y.~Su, Y.~Zhou, J.~Ren, J.~Zhang, and A.~Corigliano, ``Design,
  fabrication and experimental validation of a metaplate for vibration
  isolation in {MEMS},'' \emph{Journal of Microelectromechanical Systems},
  vol.~29, no.~5, pp. 1401--1410, 2020.

\bibitem{Zega2022}
V.~Zega, C.~Gazzola, A.~Buffoli, M.~Conti, L.~G. Falorni, G.~Langfelder, and
  A.~Frangi, ``A defect-based mems phononic crystal slab waveguide,'' in
  \emph{2022 IEEE 35th International Conference on Micro Electro Mechanical
  Systems Conference (MEMS)}, 2022, pp. 176--179.

\bibitem{Cassella2022DelayLine_a}
O.~Kaya, X.~Zhao, and C.~Cassella, ``Frequency reprogrammable {Al0.7Sc0.3N}
  acoustic delay line with up to 13.5 \% bandwidth,'' in \emph{2022 Joint
  Conference of the European Frequency and Time Forum and IEEE International
  Frequency Control Symposium (EFTF/IFCS)}, 2022, pp. 1--4.

\bibitem{Cassella2022DelayLine_b}
O.~Kaya, X.~Zhao, and C.~Cassella, ``An aluminum scandium nitride
  ({Al0.64Sc0.36N}) two-dimensional-resonant-rods delay line with 7.5\%
  bandwidth and 1.8 {dB} loss,'' in \emph{2022 IEEE 35th International
  Conference on Micro Electro Mechanical Systems Conference (MEMS)}, 2022, pp.
  1018--1021.

\bibitem{GRINLensMicro2016}
J.~Zhao, B.~Bonello, L.~Becerra, O.~Boyko, and R.~Marchal, ``Focusing of
  rayleigh waves with gradient-index phononic crystals,'' \emph{Appl. Phys.
  Lett.}, vol. 108, pp. 1--5, 2016.

\bibitem{Colombi2016}
A.~Colombi, D.~Colquitt, P.~Roux, S.~Guenneau, and R.~V. Craster, ``A seismic
  metamaterial: The resonant metawedge,'' \emph{Sci. Rep.}, vol.~6, no.~1, pp.
  1--6, 2016.

\bibitem{DePonti2019}
J.~M. {De Ponti}, A.~Colombi, R.~Ardito, F.~Braghin, A.~Corigliano, and R.~V.
  Craster, ``{Graded elastic metasurface for enhanced energy harvesting},''
  \emph{New J. Phys.}, vol.~22, no. 013013, 2019.

\bibitem{Chaplain2020Delineating}
G.~Chaplain, D.~Pajer, J.~M. De~Ponti, and R.~Craster, ``Delineating rainbow
  reflection and trapping with applications for energy harvesting,'' \emph{New
  J. Phys.}, vol.~22, no.~6, p. 063024, 2020.

\bibitem{roto_Wag}
Y.~Chen, M.~Kadic, and M.~Wegener, ``Roton-like acoustical dispersion relations
  in {3D} metamaterials,'' \emph{Nature communications}, vol.~12, no.~1, pp.
  1--8, 2021.

\bibitem{Riva-time}
\BIBentryALTinterwordspacing
Y.~Xia, E.~Riva, M.~I.~N. Rosa, G.~Cazzulani, A.~Erturk, F.~Braghin, and
  M.~Ruzzene, ``Experimental observation of temporal pumping in
  electromechanical waveguides,'' \emph{Phys. Rev. Lett.}, vol. 126, p. 095501,
  Mar 2021. [Online]. Available:
  \url{https://link.aps.org/doi/10.1103/PhysRevLett.126.095501}
\BIBentrySTDinterwordspacing

\bibitem{DePonti2021Selective}
J.~M. De~Ponti, L.~Iorio, E.~Riva, R.~Ardito, F.~Braghin, and A.~Corigliano,
  ``Selective mode conversion and rainbow trapping via graded elastic
  waveguides,'' \emph{Phys. Rev. Appl.}, vol.~16, no.~3, p. 034028, 2021.

\bibitem{Pipe2022}
\BIBentryALTinterwordspacing
G.~J. Chaplain and J.~M. De~Ponti, ``The elastic spiral phase pipe,''
  \emph{Journal of Sound and Vibration}, vol. 523, pp. 1--15, Apr 2022.
  [Online]. Available:
  \url{https://www.sciencedirect.com/science/article/pii/S0022460X21007185}
\BIBentrySTDinterwordspacing

\bibitem{Khanikaev2015}
A.~B. Khanikaev, R.~Fleury, S.~H. Mousavi, , and A.~Alu, ``Topologically robust
  sound propagation in an angular-momentum-biased graphene-like resonator
  lattice,'' \emph{Nat. Commun.}, vol.~6, no.~1, pp. 1--7, 2015.

\bibitem{Mousavi2015}
S.~H. Mousavi, A.~B. Khanikaev, and Z.~Wang, ``Topologically protected elastic
  waves in phononic metamaterials,'' \emph{Nat. Commun.}, vol.~6, pp. 1--5,
  2015.

\bibitem{Huber2016}
S.~D. Huber, ``Topological mechanics,'' \emph{Nat. Phys.}, vol.~12, p.
  621–623, 2016.

\bibitem{ChaplainTopological2020}
G.~Chaplain, J.~M. De~Ponti, G.~Aguzzi, A.~Colombi, and R.~V. Craster,
  ``Topological rainbow trapping for elastic energy harvesting in graded
  {S}u-{S}chrieffer-{H}eeger systems,'' \emph{Phys. Rev. Applied}, vol.~14, p.
  054035, Nov 2020.

\bibitem{PhysRevApplied.19.034079}
\BIBentryALTinterwordspacing
J.~M. De~Ponti, L.~Iorio, G.~J. Chaplain, A.~Corigliano, R.~V. Craster, and
  R.~Ardito, ``Tailored topological edge waves via chiral hierarchical
  metamaterials,'' \emph{Phys. Rev. Appl.}, vol.~19, p. 034079, Mar 2023.
  [Online]. Available:
  \url{https://link.aps.org/doi/10.1103/PhysRevApplied.19.034079}
\BIBentrySTDinterwordspacing

\bibitem{DePonti2021b}
J.~M. {De Ponti}, L.~Iorio, E.~Riva, F.~Braghin, A.~Corigliano, and R.~Ardito,
  ``Enhanced energy harvesting of flexural waves in elastic beams by bending
  mode of graded resonators,'' \emph{Front. Mater.}, vol.~8, pp. 1--7, 2021.

\bibitem{Tsakmakidis2007}
K.~L. Tsakmakidis, A.~D. Boardman, and O.~Hess, ``{‘Trapped rainbow' storage
  of light in metamaterials},'' \emph{Nature}, vol. 450, no. 7168, pp.
  397--401, nov 2007.

\bibitem{Zhu2013}
J.~Zhu, Y.~Chen, X.~Zhu, F.~J. Garcia-Vidal, X.~Yin, W.~Zhang, and X.~Zhang,
  ``Acoustic rainbow trapping,'' \emph{Scientific Reports}, vol.~3, no. 1728,
  pp. 1--6, 2013.

\bibitem{DePonti2021Book}
J.~M. De~Ponti, \emph{Graded Elastic Metamaterials for Energy
  Harvesting}.\hskip 1em plus 0.5em minus 0.4em\relax Cham: Springer
  International Publishing, 2021.

\bibitem{Zhao2022}
\BIBentryALTinterwordspacing
B.~Zhao, H.~R. Thomsen, J.~M. {De Ponti}, E.~Riva, B.~{Van Damme},
  A.~Bergamini, E.~Chatzi, and A.~Colombi, ``A graded metamaterial for
  broadband and high-capability piezoelectric energy harvesting,'' \emph{Energy
  Conversion and Management}, vol. 269, p. 116056, 2022. [Online]. Available:
  \url{https://www.sciencedirect.com/science/article/pii/S0196890422008445}
\BIBentrySTDinterwordspacing

\bibitem{laermer_method_1996}
\BIBentryALTinterwordspacing
F.~Laermer and A.~Schilp, ``Method of anisotropically etching silicon,'' patent
  {US} 5\,501\,893A. [Online]. Available:
  \url{https://patents.google.com/patent/US5501893A/en}
\BIBentrySTDinterwordspacing

\bibitem{art:Rajagopal12}
P.~Rajagopal, M.~Drozdz, E.~A. Skelton, M.~J. Lowe, and R.~V. Craster, ``On the
  use of absorbing layers to simulate the propagation of elastic waves in
  unbounded isotropic media using commercially available finite element
  packages,'' \emph{NDT \& E International}, vol.~51, pp. 30--40, 2012.

\bibitem{Achenbach1973}
\BIBentryALTinterwordspacing
J.~D. Achenbach, \emph{Wave Propagation in Elastic Solids}, 1973. [Online].
  Available:
  \url{https://www.sciencedirect.com/book/9780720403251/wave-propagation-in-elastic-solids}
\BIBentrySTDinterwordspacing

\bibitem{Graff1991}
\BIBentryALTinterwordspacing
K.~Graff, \emph{Wave Motion in Elastic Solids}, ser. Dover Books on Physics
  Series.\hskip 1em plus 0.5em minus 0.4em\relax Dover Publications, 1991.
  [Online]. Available: \url{https://books.google.it/books?id=5cZFRwLuhdQC}
\BIBentrySTDinterwordspacing

\end{thebibliography}


\end{document}